\documentclass{elsart3p}

\usepackage{graphicx}

\usepackage{amssymb}
\usepackage{multirow}

\begin{document}

\begin{frontmatter}

\title{Status Report and Future Prospects on LUNASKA Lunar Observations with ATCA}

\author[1]{C.W. James}
\ead{clancy.james@adelaide.edu.au}
\author[2]{R.D. Ekers}
\author[3]{J. Alvarez-Mu\~niz}
\author[1]{R.J. Protheroe}
\author[4]{R.A. McFadden}
\author[2]{C.J. Phillips}
\author[2]{P. Roberts}

\address[1]{Dept. of Physics, School of Chemistry \& Physics, Univ.\ of Adelaide, SA 5005, AUSTRALIA}
\address[2]{Australia Telescope National Facility, Epping, NSW 1710, AUSTRALIA}
\address[3]{Dept. F\'\i sica de Part\'\i culas \& IGFAE, Univ.\ Santiago de Compostela, 15782 Santiago, SPAIN}
\address[4]{School of Physics, Univ.\ of Melbourne, VIC 3010,AUSTRALIA}

\begin{abstract}
LUNASKA (Lunar UHE Neutrino Astrophysics with the Square Kilometre Array) is a
theoretical and experimental project developing the lunar Cherenkov technique for the
next generation of giant radio-telescope arrays. Here we report on a series of
observations with ATCA (the Australia Telescope Compact Array). Our current
observations use three of the six 22m ATCA antennas with a 600 MHz bandwidth at
1.2-1.8 GHz, analogue dedispersion filters to correct for the typical night-time
ionospheric dispersion, and state-of-the-art 2 GHz FPGA-based digital pulse detection
hardware. We have observed so as to maximise the UHE neutrino sensitivity in the region surrounding
the galactic centre and to Centaurus A, to which current limits on the highest-energy
neutrinos are relatively weak.
\end{abstract}

\end{frontmatter}

\section{Introduction}
\label{intro}

The Lunar Cherenkov (LC) technique, in which radio-telescopes
search for pulses of microwave-radio radiation produced via
the Askaryan effect \cite{askaryan62} from UHE particle interactions in the Lunar regolith, is a
promising method for detecting the highest energy cosmic rays (CR)
and neutrinos. Proposed by Dagkesamanskii
and Zheleznykh~\cite{Dagkesamanskii} and first attempted by Hankins, Ekers \&
O'Sullivan~\cite{Parkes96} using the Parkes radio telescope, subsequent
experiments at Goldstone (GLUE) \cite{Gorham04}, Kalyazin
\cite{Beresnyak05}, and Westerbork \cite{Scholten08}
have placed limits on an isotropic flux of UHE
neutrinos.

The Square Kilometre Array (SKA; \cite{SKAwebsite}),
a giant radio array of total collecting area $1$ km$^2$
to be completed by $\sim$2020, will offer unprecedented
sensitivity, and have the potential to observe both
a cosmogenic neutrino flux from GZK interactions of
UHE CR and the UHE CR themselves \cite{Ekers08}.
One aim of the LUNASKA project (Lunar UHE Neutrino Astrophysics with the SKA)
is to develop experimental methods scalable to giant, broad-bandwidth radio arrays
such as the SKA. For this purpose, we have been using the Australia Telescope Compact Array (ATCA), a
radio interferometer of six $22$-m dishes along a $6$~km E-W baseline
located in New South Wales, Australia. Here we report on our techniques,
which have allowed us to achieve a lower detection threshold than
other LC experiments, and have the greatest exposure to $10^{21}-3\times10^{22}$~eV
neutrinos coming from the vicinity of Centaurus A and the Galactic centre over all
detection experiments.

\section{Experimental Set-Up}
\label{setup}

\begin{figure}
\centering
\includegraphics[width=6cm]{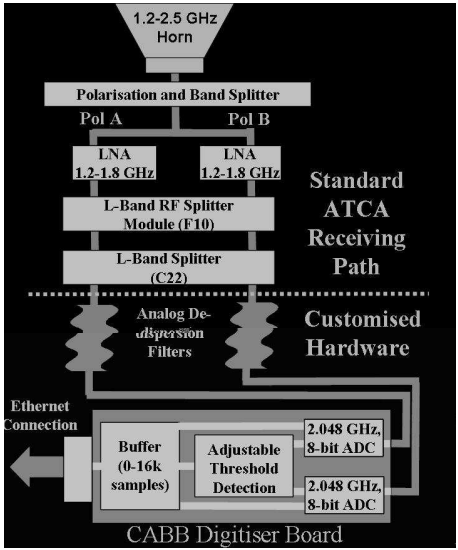}
\caption{Diagram of the signal path at each antenna. Our customised hardware is described
in text.}
\label{block_diagram}
\end{figure}

We have implemented the hardware described below on three of the six ATCA antennas,
CA01, CA03, and CA05, with a maximum baseline of $750$~m. In each, we installed
specialised pulse de-dispersion and
detection hardware, with the full signal path at each antenna shown in
Fig.\ \ref{block_diagram}. We have been utilising an FPGA-based back-end,
the ``CABB digitiser board'', installed as part of the on-going
Compact Array BroadBand upgrade, which allows us to process the full
$600$~MHz ($1.2$-$1.8$~GHz) bandwidth provided by the standard ATCA L-band signal path.
This is split into dual linear polarisations, passed through an analogue de-dispersion
filter to correct for the effects of the Earth's ionosphere,
and then $8$-bit-sampled at $2.048$-GHz. A simple threshold trigger algorithm is then
applied which sends back a $256$~sample ($125$~ns) buffer of both polarisations
to the control room --- along with antenna-specific clock times accurate to $0.5$~ns ---
should the voltage on either polarisation exceed an adjustable threshold.

To coherently dedisperse our full $600$~MHz bandwidth and recover our full sensitivity,
we have made use of innovative new analogue dedispersion filters. The filters consist of $1$~m of tapered microwave waveguide wrapped for easy of
storage into a spiral pattern, with the continuous sum
of reflections along their length producing a frequency-dependent delay varying
contrary to (and thus correcting for) the delay induced by the Earth's ionosphere. While the filters
can only correct for a fixed delay, the ionosphere over the
antenna during night-time hours (at least near solar cycle minimum) is
relatively stable, producing a typical $4$~ns of dispersion across our bandwidth at zenith.
Therefore we set the filters to correct for $5$~ns of dispersion, i.e.\ the expected value
when the Moon is at $53^{\circ}$ elevation.

For the observations reported here, the full continuous bandwidth could
not be returned to a central location
for coincident triggering, though this will not be the case for future observations.
Hence our use of only three antennas --- until we combine information from
all antennas in real-time, we are limited by the sensitivity of each,
which we can only partially recover with by increasing our trigger rate. 
We currently set the thresholds so that each
polarisation channel would be triggering at $\sim 30$~Hz of a maximum $3000$~Hz, for an effective dead-time on a three-fold trigger of approximately $6$\%.

\section{Observations}
\label{obs}

A summary of our observation runs is given in Tbl.\ \ref{obstbl}, covering a trial period in May 2007, and our main observing runs in February and May 2008. The 2007 and February 2008 runs were tailored to `target' a broad ($\gtrsim 20^{\circ}$) region of the sky near the galactic centre, harbouring the closest supermassive black hole and potential accelerator of UHE CR.
Therefore for these runs we pointed the antenna towards the lunar centre, since this mode maximises coverage of the lunar limb (from which we expect to see the majority of pulses) and hence we achieve the greatest total effective aperture.
Our May 2008 observing period targeted Centaurus A only,
a nearby active galaxy which could potentially account for two of the UHE CR events observed by the Pierre Auger observatory \cite{AugerScience07}. Regardless of their source, this suggests the likelihood of an accompanying excess of UHE neutrinos, and we do not exclude the possibility of seeing the UHE CR themselves.
We therefore pointed the antenna at that part of the lunar limb closest to Cen A in order to maximise sensitivity to UHE particles from this region \cite{LUNASKA_theory}.

\begin{table}
\begin{center}
\begin{tabular}{|l||c || c c c || c c c|}
\hline
 & May '07 & \multicolumn{3}{c||}{Feb '08} & \multicolumn{3}{c|}{May '08} \\
\hline
\hline
Day	&	$7^{th}$ & $26^{th}$ & $27^{th}$ & $28^{th}$ & $17^{th}$ & $18^{th}$ & $19^{th}$ \\
$t_{\rm obs}$ (mins) & 210	& 275	& 330	& 255	& 320	& 370	& 435 \\
\hline
\end{tabular}
\caption{Observation dates (nights thereof) and total time $t_{\rm obs}$ spent observing the Moon in detection mode of our observing runs.}
\label{obstbl}
\end{center}
\end{table}

For each period, we observed nominally between the hours of $10$~pm and $6$~am local time. Since we were utilising new equipment in the midst of an upgrade, we did not have ns timing available for our 2007 observations. In 2008 we had to align the clocks manually by observing the bright quasar 3C273 and correlating the emission between antennas. Since $T_{\rm sys}$ measurements were available over only part of our bandwidth, we used the Moon as our absolute sensitivity calibrator
by setting trigger levels to zero to collect an unbiased sample of data when pointing both on and off the Moon.

\section{Analysis}
\label{analysis}

Our off-line processing involves searching through the candidate triggered
events for three-fold coincidences with timing, duration, dispersion, and polarisation consistent with
coherent Cherenkov pulses arriving from the direction of the Moon. We have found
the timing constraint to be both the strongest, most reliable, and most easily automated,
with the remaining events typically few enough to sort manually.

\begin{figure}
\centering
\includegraphics[width=7cm]{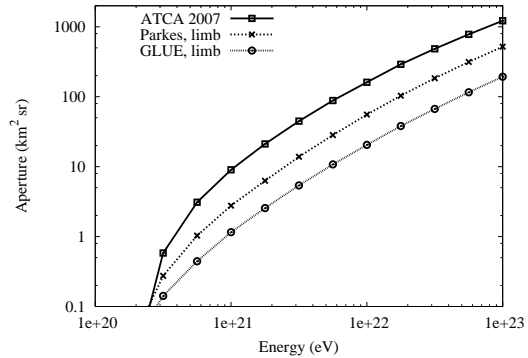}
\caption{
Effective aperture of the (inferior) $2007$ LUNASKA lunar observations under the assumption of no positive detections. Also shown are our calculations \cite{JP} for the Parkes experiment and GLUE.}
\label{applims}
\end{figure}

For our $2007$ observations, the analysis is complete. Since we did not have accurate clocks at each antenna, we had to utilise PC clocks with $\pm 1$~ms accuracy only, and thus did not have enough discriminatory power to accept candidate events as having lunar origin. However, our rejection power was still great, and out of $150,000$ candidates for each antenna we were left with only $4$ with consistent polarisation, arrival times simultaneous to within $1$~ms, and having pulse-like structure. Since we expected $\sim 6$ random events from purely thermal noise to fulfil these criteria we do not have reason to suspect any of these candidates to be of real lunar origin. Fig.\ \ref{applims} plots our calculations of the effective aperture to UHE $\nu$ calculated for this run, assuming a correct dedispersion. The broad bandwidth of our observations compensates for our lower collecting area when compared with previous experiments at Parkes and Goldstone. Since we see the entire Lunar limb, our effective aperture is greater than previous lunar Cherenkov experiments (as discussed in detail by \cite{Ekers08}, except in the $E_{\nu} > 3 \times 10^{23}$~eV range where NuMoon dominates \cite{Scholten08}.

\begin{figure}
\centering
\includegraphics[width=7cm]{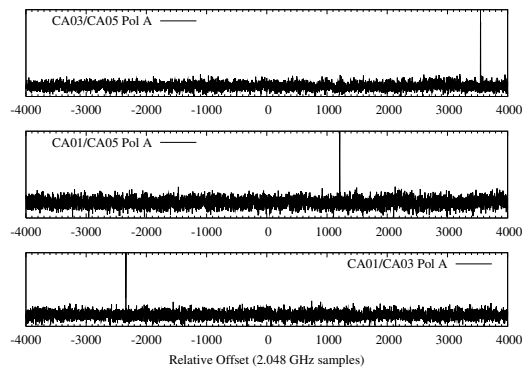}
\caption{
The sum of $200$ cross-correlations of buffers from antennas CA01, CA03, and CA05, triggered near-simultaneously while pointing at the bright quasar 3C273, and corrected for the quasar position.
}
\label{timing}
\end{figure}

For the $2008$ observations, we have of order $6 \times 10^{6}$ candidate events for each antenna, and the analysis is not yet complete (all calculations shown here are based on the sensitivity of our inferior 2007 observations).
Except during periods of intense RFI, we can readily cross-correlate the buffers to achieve timing to $1$~ns accuracy, an example of which is shown in Fig.\ \ref{timing}. Assuming the candidates are all random thermal events, our timing gives a false detection probability over our entire experiment of less than $10^{-5}$. A preliminary analysis indicates that a large fraction are due to terrestrial RFI. Although to a $1$-D baseline some ground-based RFI received through an antenna side-lobe may coincidentally appear to come from the Moon, these can be eliminated on the dispersion, polarisation and duration criteria.

\begin{figure*}
\centering
\includegraphics[width=12cm]{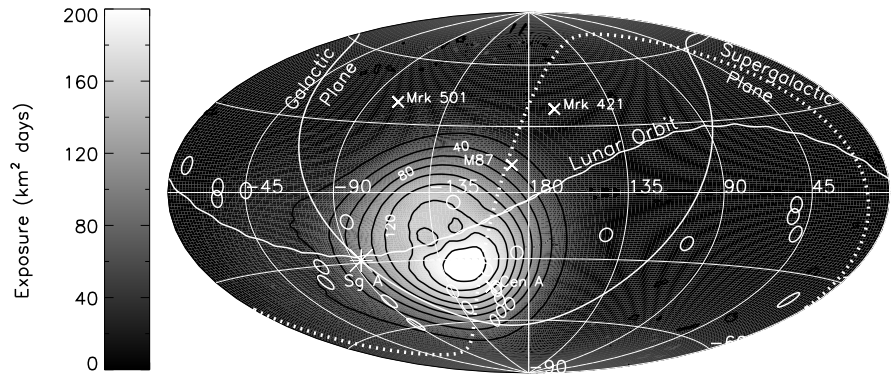}
\caption{Exposure map ($20,40,\dots,180$ km$^2$-day contours) in celestial coordinates, centred at $(\alpha,\delta)=(180^{\circ},0^{\circ})$, of our ATCA lunar observations to $10^{22}$~eV neutrinos, assuming the sensitivity of our 2007 observations. See \cite{JP}.}
\label{exposure}
\end{figure*}

While our expected limit on an isotropic flux of UHE $\nu$ will not be competitive with that from ANITA, our goal was to target specific regions of the sky. Using preliminary calculations based off our $2007$
sensitivity, Fig.\ \ref{exposure} shows an exposure map for our observations (see \cite{LUNASKA_theory}), while Tbl.\ \ref{exposuretbl} shows the improvement in exposure to Cen A and Sgr A* over that of Parkes and GLUE.

\begin{table}
\begin{center}
\begin{tabular}{|l| c c c | c c c|}
\hline
	& \multicolumn{3}{c|}{Parkes \& GLUE} & \multicolumn{3}{c|}{LUNASKA} \\
\hline
Energy (eV) & $10^{21}$ & $10^{22}$ & $10^{23}$ & $10^{21}$ & $10^{22}$ & $10^{23}$ \\
\hline
Sgr A*	& 0.5	& 14	& 175	& 3.7	& 75	& 565	\\
Cen A	& 0.015	& 2.1	& 43	& 9.3	& 145	& 970 \\
\hline
\end{tabular}
\caption{Our accumulated exposure (km$^2$-days) to Cen A and Sag A from the Parkes experiment and GLUE (our calculations), and for our LUNASKA observations with ATCA, based on the inferior $2007$ sensitivity. See \cite{JP}.}
\label{exposuretbl}
\end{center}
\end{table}

\section{Conclusions}
\label{conclusions}

Our LUNASKA lunar observations are continuously improving, both as new hardware
becomes available, and as we become more experienced at observing in a ns radio
environment. We have already achieved the greatest sensitivity for a Lunar
Cherenkov experiment, and accumulated the greatest exposure to UHE neutrinos
in the $10^{21}-10^{23}$~eV range from the vicinity of Sgr A* and Cen A. We have
also demonstrated that an array of antennas observing
over a broad bandwidth is extremely efficient for disciminating between terrestrial
RFI and true Lunar pulses. The next stage is to implement
real-time coincidence logic between antennas and improve the RFI filtering, so we
expect our $2009$ observations --- for which time has been allocated
--- to be the most sensitive yet.

\section{Acknowledgments}
This research was supported by the Australian
Research Council's Discovery Project funding scheme (project
numbers DP0559991 and DP0881006).
The Australia Telescope
Compact Array is part of the Australia Telescope which is funded by the
Commonwealth of Australia for operation as a National
Facility managed by CSIRO. J.A-M thanks Xunta de Galicia
(PGIDIT 06 PXIB 206184 PR) and
Conseller\'\i a de Educaci\'on (Grupos de Referencia Competitivos --
Consolider Xunta de Galicia 2006/51).

\end{document}